\documentclass{mem}
\usepackage{natbib}\usepackage{txfonts}\usepackage{balance}
\usepackage{graphicx}
\usepackage[a4paper]{hyperref}
\idline{0}{1}
\begin{document}
\def\teff{$T\rm_{eff }$}
\def\kms{$\mathrm {km s}^{-1}$}

\title{
Metals and dust in high redshift AGNs
}

   \subtitle{}

\author{
R.~Maiolino\inst{1}, 
T.~Nagao\inst{1,2}, A.~Marconi\inst{1}, R.~Schneider\inst{1},
S.~Bianchi\inst{3},
M.~Pedani\inst{4}, A.~Pipino\inst{5}, F.~Matteucci\inst{5},
P.~Cox\inst{6} \and P.~Caselli\inst{1}
          }

  \offprints{R. Maiolino}

\institute{
INAF --
Osservatorio Astrofisico di Arcetri, Largo E. Fermi 5,
I-50125 Firenze, Italy
\and
National Astronomical Observ. of Japan, 2-21-1 Osawa, Mitaka,
Tokyo 151-8588, Japan
\and
INAF -- IRA (sez. Firenze), Largo E. Fermi 5,
I-50125 Firenze, Italy
\and
INAF -- Telescopio Naz. Galileo, P.O. Box 565, E-38700 Santa Cruz de la Palma,
Spain
\and
Dip. di Astronomia, Universit\`a di Trieste, via G.B.~Tiepolo 11, I-34127,
 Trieste Italy
\and
IRAM, 300 rue de la Piscine, 38406 St.-Marin-d'H\'eres, France
}

\authorrunning{Maiolino et al.}

\titlerunning{Metals and dust at high-z}

\abstract{
We summarize some recent results on the metallicity and dust properties
of Active Galactic Nuclei (AGN) at high redshift (1$<$z$<$6.4).
By using the spectra of more than 5000 QSOs from the SDSS we find no evidence 
for any
metallicity evolution in the redshift range 2$<$z$<$4.5, while there is a significant
luminosity--metallicity dependence. These results are confirmed by the
spectra of a smaller sample of narrow line AGNs at high-z (QSO2s and radio galaxies).
The lack of metallicity evolution is interpreted both as a consequence of the cosmic
downsizing (QSO massive hosts evolve rapidly and at high redshift)
and as a selection effect resulting from the
joint QSO-galaxy evolution (most QSOs are observable at the late evolutionary
stages of their hosts).
The luminosity--metallicity relation is interpreted as a
consequence of the mass--metallicity relation in the host galaxies of QSOs, but
a relationship with the accretion rate is also possible.
The lack of metallicity evolution is observed even in the
spectra of the most distant QSOs known (z$\sim$6). This result is particularly
surprising for elements such as Fe, C and Si, which are subject to a delayed
enrichment, and requires that the hosts of these QSOs formed in short
bursts and
at very high redshift (z$>$10). The properties of dust in high-z QSOs are discussed
within the context of the dust production mechanisms in the early universe.
The dust
extinction curve is observed to evolve beyond z$>$4, and by z$\sim$6 it is well
described by the properties expected for dust produced by SNe, suggesting that the
latter is the main mechanism of dust production in the early universe.
We also show that the huge dust masses observed in distant QSOs can be accounted for
by SN dust within the observational constraints currently available. Finally, we
show that QSO winds, which have been proposed as an alternative mechanism of dust
production, may also contribute significantly to the total
dust budget at high redshift.

}
\maketitle{}

\begin{figure*}[t!]
\centerline{
\resizebox{10truecm}{!}{\includegraphics[clip=true]{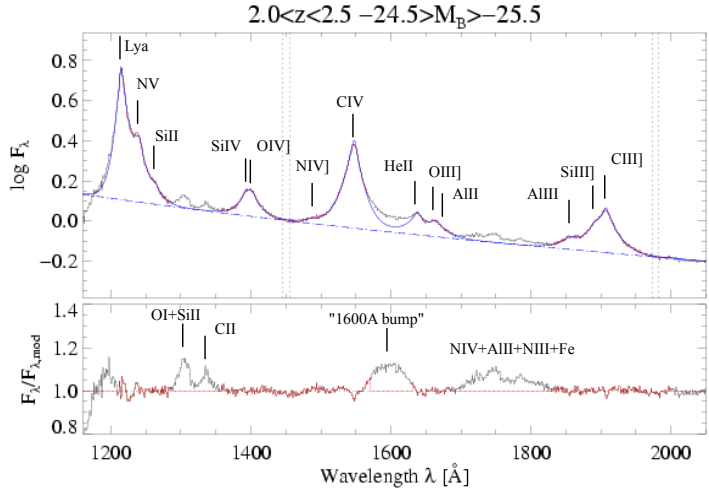}}
}
\caption{\footnotesize
One of the QSO composite spectra obtained by stacking a few hundreds SDSS spectra.
In the upper panel the continuous smooth line shows the best fit to the main spectral
lines. The bottom panel shows
the residuals of the fit with the identification of additional
spectral lines.
}
\label{qso_comp}
\end{figure*}

\section{Introduction}

Quasars and Active Galactic Nuclei (AGNs)
in general are powerful tools to investigate the
properties of the interstellar medium (ISM) in distant galaxies.
Indeed, the huge
luminosities characterizing such galactic nuclei, excite large masses
of gas in their host galaxies, which as a consequence emit strong atomic
(and molecular) lines
that can be observed even in very distant systems. These lines
can be used to study in
detail the ISM properties, such as metal abundances. Moreover, the strong
nuclear continuum can be used to investigate in detail reddening effects and,
therefore, study the dust properties. Since large samples of AGNs are now
available over large redshift intervals, they can be used to trace the evolution
of the ISM as a function of the cosmic age. Since metals and
dust are products of the stellar evolution, the ISM investigation through AGNs
can be regarded as a powerful tool to constrain the scenarios of galaxy evolution.
In this paper we summarize some recent results obtained by us on the
metal abundances and dust properties of AGNs spanning a wide redshift
range (1$<$z$<$6.4), including some preliminary results based on ongoing work.

\section{The metallicity of the BLR at 2$<$z$<$4.5}

We have used more than 5000 QSO spectra from the Sloan Digital Sky Survey
(SDSS) data release 2
(DR2) to investigate the metallicity of the Broad Line Region (BLR)
across the redshift
range 2$<$z$<$4.5 and over the luminosity range $\rm -24.5 < M_B < -29.5$
\citep{nagao06a}. The
huge number of objects allow us to break the degeneracy between redshift and
luminosity dependence of the metallicity, which plagued most of the previous
studies. In particular, the number of objects is large enough that within each
redshift bin we can investigate the metallicity-luminosity dependence and,
viceversa, within each luminosity bin we can study the metallicity-redshift
dependence. To improve the signal-to-noise ratio we have stacked the spectra within
each redshift-luminosity bin, resulting in a grid of 22 high quality,
composite spectra. Fig.~\ref{qso_comp}
shows one of these composite spectra with a wealth of
broad emission lines. In each spectrum the emission lines were deblended by
using an algorithm described in \cite{nagao06a} and delivering accurate
fluxes for individual lines.
A clear, empirical result is that most line ratios depend significantly
on luminosity, but do {\it not} vary significantly with redshift.

\begin{figure}[t!]
\resizebox{\hsize}{!}{\includegraphics[clip=true]{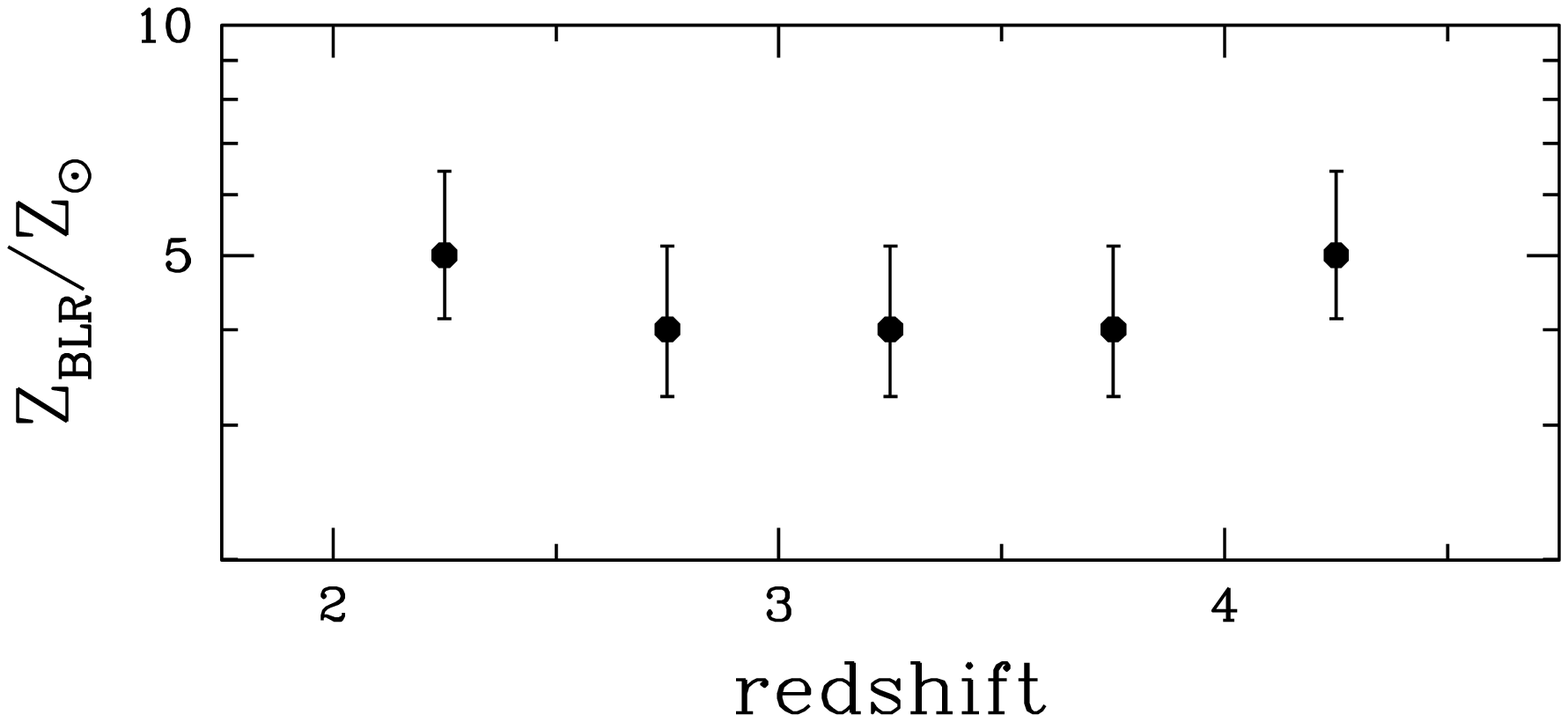}}
\vskip0.2truecm
\resizebox{\hsize}{!}{\includegraphics[clip=true]{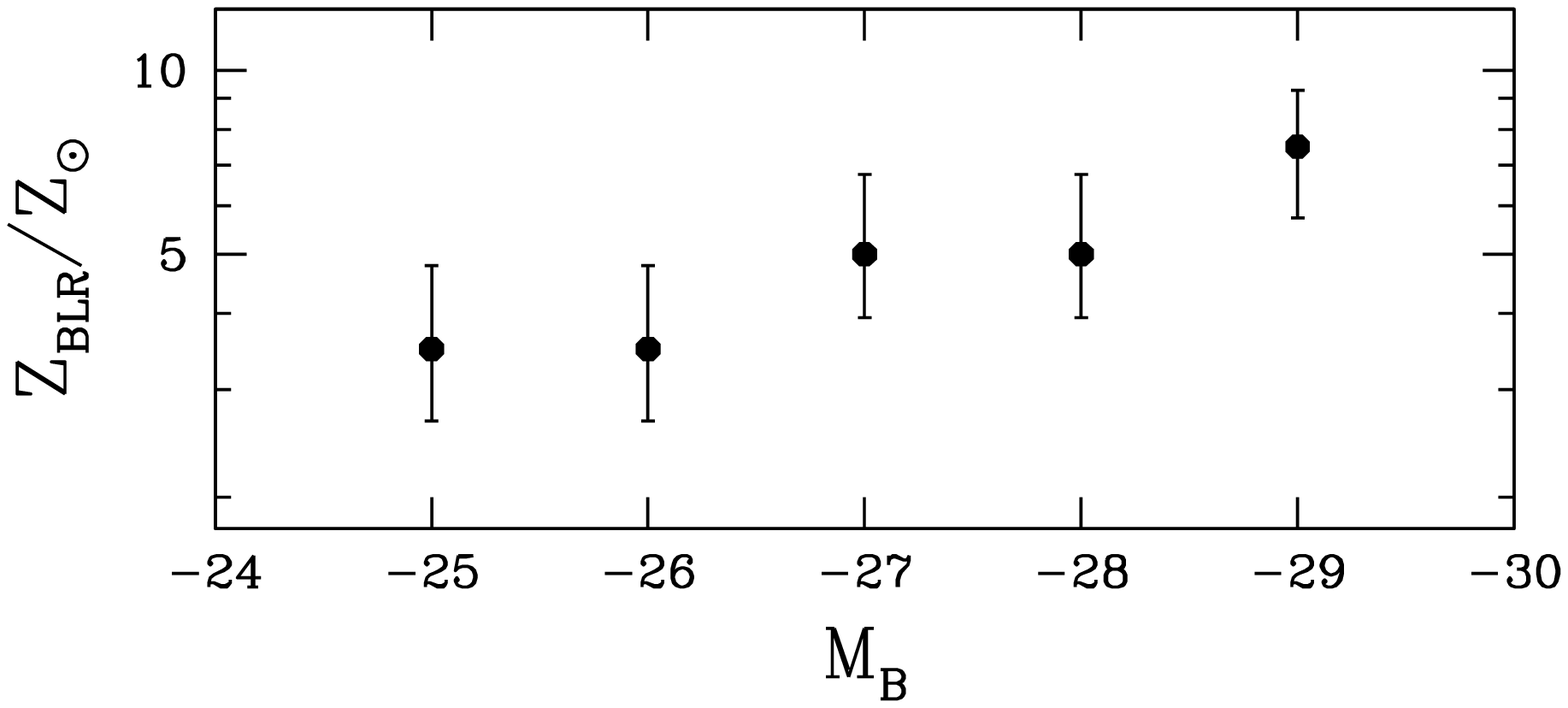}}
\caption{\footnotesize
Average metallicity as a function of redshift (upper panel) and average
metallicity as a function of luminosity (lower panel). No evidence is found for
a metallicity evolution in redshift, while there is a significant
luminosity-metallicity dependence.
}
\label{Z_trends}
\end{figure}

The several line ratios available for each composite spectra were compared
with detailed photoionization models, presented in \cite{nagao06a},
to infer the metallicity of the BLR (here abundances are assumed to scale
proportionally to solar, except for nitrogen that is assumed to scale
quadratically).
The first important result is that
the metallicities turn out to be super-solar in all cases.
The trends as a function of redshift and luminosity are summarized
in Fig.~\ref{Z_trends},
which shows the metallicities averaged in redshift bins (top panel)
and in luminosity bins (bottom panel). The striking result from the top panel
of Fig.~\ref{Z_trends}
is that the metallicities in the nuclear region of these quasars
are insensitive to redshift, although the QSO sample
spans about 90\% of the age of the universe. The additional result is that
the average metallicity is an increasing function
of the QSO luminosity. The latter result can be interpreted as a consequence
of the mass-metallicity relation in galaxies
\citep[e.g.][]{tremonti04}, due to the tendency of more massive galaxies
to retain metals more effectively. Indeed, if the QSO luminosity is
proportional to the black-hole mass (i.e. the Eddington ratio is on average
constant for QSOs with different luminosities), then the $\rm M_{BH}-M_{bulge}$
relation \citep[e.g.][]{marconi03} yields a relationship between QSO luminosity
and galaxy mass, and therefore a dependence on metallicity.
Although this scenario can qualitatively explain the
luminosity-metallicity relation in QSOs, there are problems when a
{\it quantitative} comparison is performed
with the mass-metallicity relation observed
in galaxies. Another possibility is that the luminosity--metallicity relation
actually reflects a relationship between metallicity and
accretion rate (in terms of $\rm L/L_{Edd}$), as suggested by
\cite{shemmer04}. Unfortunately we cannot test the latter scenario, since
the information on the black hole mass, and therefore on $\rm L/L_{Edd}$, was
lost in the process of stacking the spectra.

Much more puzzling is the lack of any metallicity evolution over the
wide redshift range covered by our sample. Indeed, most galaxies
at redshift larger than 2 should be undergoing (or have undergone)
strong star formation, and therefore should be characterized by
a differential chemical enrichment as a function of
redshift (specifically, higher metallicities at lower redshifts).
In the following we discuss three possible effects or scenarios that may
explain the lack of evolution for the quasars BLR.

\begin{figure}[t!]
\resizebox{\hsize}{!}{\includegraphics[clip=true]{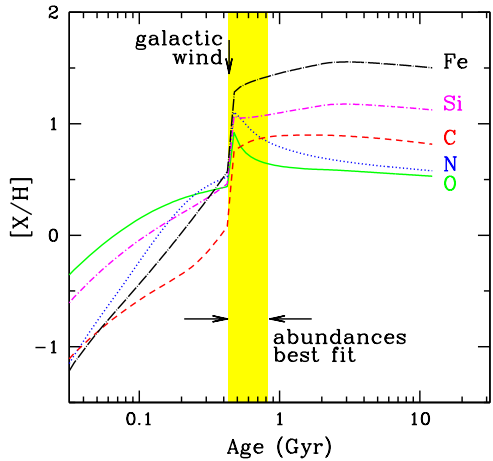}}
\caption{\footnotesize
Abundances evolution for an elliptical galaxy including feedback effects. The downward
arrow indicates the onset of the galactic wind. The shaded area indicates the abundance
sets which best fit the line ratios observed in the QSO spectra.
}
\label{abundances_ev}
\end{figure}

\subsection{Selection effects associated with the QSO-galaxy co-evolution}

The lack of metallicity evolution may be a consequence of QSOs being
selectively observed at specific epochs, and in particular when they are
already aged and chemically mature. Such an effect can be understood within the
context of the recent models of QSO-galaxy co-evolution
\citep[e.g.][]{granato04,dimatteo05}.
According to these models, during the early stages of galaxy and black hole
growth the AGN is embedded in gas and dust and it is difficult to identify it
as a QSOs, due to obscuration.
At later times the combined energy released by SNe and by the QSO wind
sweeps away large quantities of dust and gas, quenching star formation, and
allowing the direct observation of the QSO. At this stage the ISM has already
been heavily enriched. We are in the process of investigating this issue more
quantitatively (Nagao et al. in prep).
More specifically, instead of using constant abundances ratios
as discussed above, we are using results of realistic,
detailed abundances patterns based on evolutionary models including SNe
and AGN feedback. For each abundance pattern at any stage, we infer the
expected line ratios and compare them with those observed in the composite QSO
spectra. A preliminary result is shown in Fig.~\ref{abundances_ev},
where the abundances
evolution is shown for the case of a massive elliptical galaxy 
\citep[][this one
only includes feedback from SNe, a version including QSO feedback is in
preparation]{pipino04}. The discontinuity in the slope at about 0.5~Gyr is
a consequence of the onset of the galactic super-wind.
The shaded area shows the range of abundances pattern that best
match the line ratios observed in the composite spectra. The most interesting
result is that the best matches are obtained at the wind onset,
or shortly after that. On the one hand, this result nicely supports models of
QSO-galaxy co-evolution involving feedback. On the other hand,
the fact that QSOs
can be selectively (or preferentially) observed only when they are (chemically)
evolved, because their circumnuclear region has been cleared, may explain
the apparent lack of metallicity evolution in QSOs.

However, selection effects due to joint QSO-galaxy evolution cannot be the
only explanation for the observed lack of metallicity
evolution. Indeed, in Sect.~\ref{met_nlr}
we will show
that the lack of evolution applies also to narrow line radio galaxies, which
are obscured systems, hence presumably observed before the onset of the wind
(although radio galaxies are known to be hosted in massive, evolved systems).

\begin{figure*}[t!]
\centerline{
\resizebox{10truecm}{!}{\includegraphics[clip=true,angle=-90]{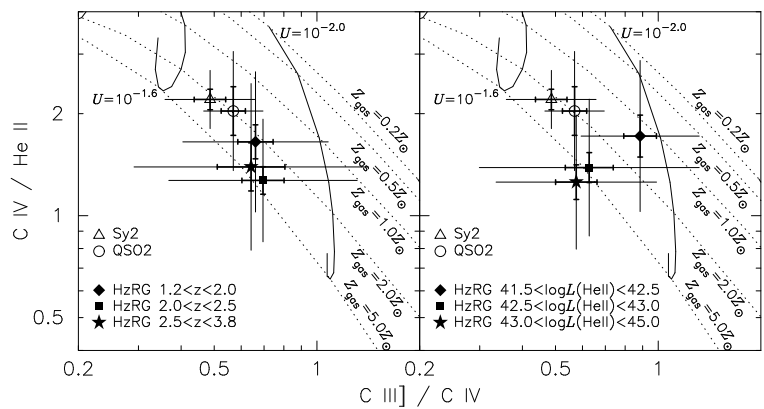}}
}
\caption{\footnotesize
Diagram showing the average line ratios $\rm CIV\lambda 1549/HeII\lambda 1640$
and $\rm CIII]\lambda 1909/CIV \lambda 1549$ for radiogalaxies grouped in bins of
redshift (left) and of HeII luminosity (right). These diagrams are a useful
diagnostic to investigate the metallicity of the Narrow Line Region. In particular
the dotted lines indicates the loci of the line ratios
at constant gas metallicity, while solid lines show loci
at constant ionization parameter. An increasing metallicity tends to decrease both line ratios.
The observed average line ratios indicate no evolution of the gas metallicity as a
function redshift, but there is a significant dependence on luminosity.
}
\label{nlr}
\end{figure*}

\subsection{Antihierachical evolution and chemical downsizing}

Recent observational studies have reported the detection of a number of high-z
massive galaxies significantly
larger than expected by classical hierarchical models
\citep[e.g.][]{fontana04}, suggesting that massive systems evolve faster
and at higher redshift than less massive galaxies. This effect is also
observed in AGN/QSO surveys, where more luminous systems (hence probably more
massive) are seen to peak their
evolution at redshift significantly higher than less luminous systems
\citep[][]{hasinger05}. This effect is known as ``antihierarchical'' growth or
``downsizing'', and it has been explained theoretically through the SN/QSO
feedback effects, which reverse the hierarchical baryonic growth with respect
to the dark matter haloes \citep{dimatteo05,granato04}. \cite{savaglio05}
found a chemical version of the downsizing phenomenon, by investigating the
mass-metallicity relation at high redshift (z$\sim$1). They found that the
M--Z relation evolves, from z=0 to z=1, very rapidly at low galactic
masses (in the sense that the metallicity decreases rapidly from z=0 to z=1),
while the chemical evolution at high masses ($\rm M_*\approx 10^{11}
M_{\odot}$) is only marginal, and probably remains constant within
uncertainties. This result indicates that more massive system evolve,
from the chemically point of view, at higher redshift and faster
than less massive systems. All the SDSS QSOs used in \cite{nagao06a} are
very luminous systems, which are probably hosted in very massive galaxies
($\rm M_* > 10^{11} M_{\odot}$) that, within the context of the downsizing
scenario, represent very extreme cases, being already fully
evolved and mature already at very high redshift. A more quantitative analysis
of the chemical downsizing applied to QSOs will be presented in a forthcoming
paper.

\subsection{BLR not representative of the host galaxy}\label{blr_not_rep}

The BLR is contained in a very small nuclear region (less than a parsec
in size),
which could be subject to a much more rapid evolution than the rest of the
galaxy. The gas mass of the BLR is small enough that a relatively
small number of SNe is enough to rapidly enrich it. Summarizing, the BLR
may not be representative of the metallicity in the host galaxy.
This issue can be tackled by estimating the metallicity of the Narrow Line
Region (NLR), which extends on sizes comparable with the host galaxy, as
discussed in the following section.

\section{The metallicity of the NLR at 1$<$z$<$4}\label{met_nlr}

We have used a sample of about 60 high-redshift narrow-line AGN
(radio galaxies and type 2 QSOs, which are both AGNs where the BLR is obscured)
to investigate the metallicity of the Narrow
Line Region (NLR) in the redshift range 1$<$z$<$4 \citep[][]{nagao06b}.
Although this is the largest
sample of narrow line AGNs available in this redshift range, it is certainly
much smaller than the huge sample available from the SDSS for broad line QSOs.
Moreover, the faintness of these objects restricts the use of their observed
spectra only to a few (narrow) emission lines. Even with these caveats, the
narrow lines probe a region much more extended than the BLR (in luminous
objects the NLR extends over several kpc), and therefore they are more suited
to probe the host galaxy. Details of this investigation are provided in
\cite{nagao06b}. Fig.~\ref{nlr} summarizes the main results through the diagram
involving the UV line ratios CIV/HeII vs. CIII]/CIV, which is a powerful
diagnostic diagram to probe the metallicity of the NLR. The dotted curves
indicate the loci at constant metallicity,
which moves from the top-right
towards the bottom-left part of the diagram with increasing metallicity.
In the left-hand side panel black symbols indicate the line ratios observed in radio
galaxies, averaged in redshift bins. The most interesting result is a lack of
evolution, in terms of line ratios (and therefore in terms of metallicity),
from z$\sim$1.6 to z$\sim$3.2, as previously found for the BLR. 
In the right-hand side panel black symbols indicate the
line ratios observed in radio galaxies, averaged in bins of HeII luminosity,
where the HeII line is assumed to be a fair indicator of the intrinsic luminosity
of these systems. These diagrams display a significant dependence
of the metallicity on luminosity, as found for the BLR. Summarizing, the
trends observed for the BLR are also observed for the NLR, suggesting that the
lack of metallicity evolution cannot be entirely ascribed to the small volume
probed by the former, nor to obscuration effects associated with evolutionary
effects.

\section{The chemical evolution of QSOs beyond z$>$5}

The SDSS have now delivered a sizeable sample of quasars even at redshift 
z$>$5, the most distant of which currently being at z=6.4 \citep[][]{fan03}.
Such high redshifts are extremely interesting to investigate within the context
of the chemical evolution. Indeed, at z$>$5 the age of the universe
is comparable with the minimum timescale required by stellar processes to enrich
the ISM and therefore, regardless of selection effects and galactic
evolutionary scenarios discussed above, some chemical evolution is expected to
occur. In particular, some elements such as iron and carbon are mostly enriched by
type Ia supernovae (SNIa) and AGB stars, which require about 1 Gyr to evolve
in classical models, while at z=6.42 the age of the universe is about 800~Myr.

\begin{figure}[t!]
\resizebox{\hsize}{!}{\includegraphics[clip=true]{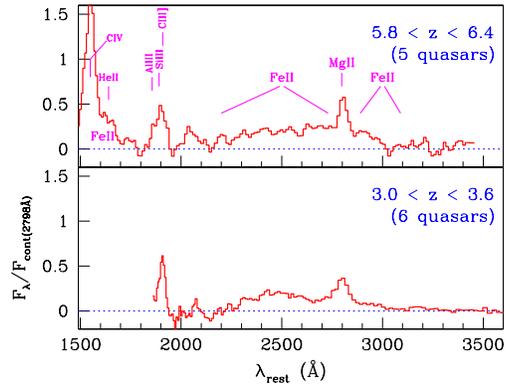}}
\caption{\footnotesize
UV rest-frame average spectra of 5 QSOs at 5.8$<$z$<$6.4 (top) showing a prominent
iron bump, similar to that observed in lower redshift QSOs (bottom), suggesting
a large abundance of Fe even at an epoch when the age of the universe was less
than 1~Gyr.
}
\label{fe}
\end{figure}

At such high redshift many of
the most prominent QSO emission lines are shifted into the
near-IR band. Our and other groups have employed deep near-IR spectroscopic
observations to investigate the metallicity and abundances in the most distant
quasars known.
The most surprising result is the evidence for
high chemical enrichment already at these early
epochs and, again, the lack of any evolution
with respect to QSOs at lower redshift. Fig.~\ref{fe}
shows the (continuum-subtracted)
average spectrum of 5 QSOs at 5.8$<$z$<$6.4 in the UV rest frame
\citep[from ][]{maiolino03}. The prominent bump at $\rm 2200<\lambda _{rest}
<3000 \AA$, produced by Fe multiplets, suggests a large abundance of iron 
(probably super-solar) in these objects. The quantitative determination
of the iron abundance through these spectra is not trivial and highly uncertain
\citep[][]{verner04}. However,
the similarity of the Fe-bump--to--MgII$\lambda 2798$
ratio between intermediate redshift and z$\sim$6 QSOs (Fig.~\ref{fe})
strongly suggests little or no evolution in terms of
iron abundance relative to the $\alpha$-elements (which are expected to be
produced on much shorter time scales by SNIIe).
Similar results were obtained by several other authors
\citep[][]{iwamuro04,freudling03,dietrich03,barth03}.
The analysis of other (broad) line ratios involving other important
elements (C, N, Si) have also
provided no clear evidence for any chemical evolution
up to z$\sim$6 \citep[][Maiolino et al. in prep.]{pentericci02}.

\begin{figure}[t!]
\resizebox{\hsize}{!}{\includegraphics[clip=true]{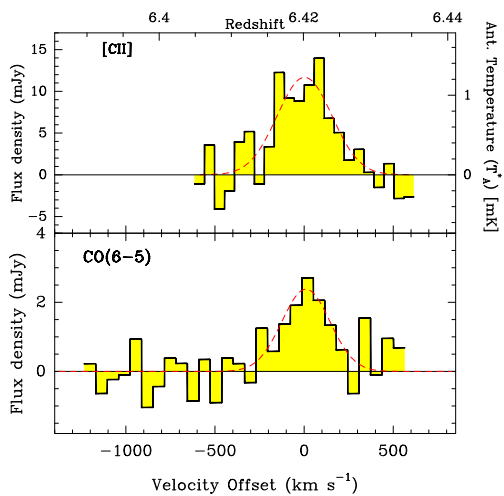}}
\caption{\footnotesize
[CII]158$\mu$m (top) and CO(5--6) (bottom) emission
lines detected in the host galaxy of the
most distant QSO SDSSJ1148+52 at z=6.4 \citep[][]{maiolino04}.
}
\label{cii}
\end{figure}

The criticism, as discussed in \ref{blr_not_rep},
that these broad lines used originate from a small region
and that may not be representative of the host
galaxy, is addressed by recent millimetric and radio observations.
Indeed, \cite{walter03}
and \cite{bertoldi03a} reported the detection of CO emission lines in the most
distant QSO SDSSJ1148+52 at z=6.4. This line is certainly emitted in the host galaxy of
the QSO and, beside tracing a large quantity of molecular gas, suggests the
presence of large amounts of carbon. However,
CO lines are optically thick and may be
intense even if the carbon abundance is reduced. A more decisive
observation was
the detection of the [CII]158$\mu$m in the same object
\citep[Fig.~\ref{cii},][]{maiolino05}. In contrast to CO,
this [CII] line is optically thin, and its intensity indicate a strong carbon
enrichment in the host galaxy of SDSSJ1148+52 at z=6.4. However,
more observations are required to quantitatively
determine the carbon abundance, and in particular the C/O ratio, as discussed
in \cite{maiolino05}.

Altogether these results on the metal enrichment of the most distant QSOs
are surprising, but not yet in striking contrast with
models of chemical evolution. Indeed, in the extreme scenario of a very rapid and
very efficient formation of massive elliptical galaxies (and occurring at very
high redshift, z$>$10), recent models can account for a strong enrichment of
critical elements, such as Fe, C, N, Si, on time scales as short as a few 100 Myr
\citep[][see also Fig.~\ref{abundances_ev}]{matteucci01,pipino04,venkatesan04}.
Within this
context, it is important to note the recent finding by \cite{mannucci05} and
\cite{mannucci06} that a population of ``fast'' SNIa can enrich the ISM on time
scales shorter than 100~Myr.

\section{Dust evolution at high redshift}

Dust, which plays a crucial role in the formation of galaxies and on their
observability, is also affected by evolutionary issues and time scales
associated with stellar processes.
More specifically, according to the classical scenario
dust is mostly formed in the envelopes of AGB stars (and late giant in general)
which take about 1~Gyr to evolve \citep[][]{morgan03,marchenko06}. As a consequence,
little dust is expected to be present at z$>$5, when the universe was younger than
1~Gyr. However, a recent unexpected and puzzling result, obtained through deep
mm and submm observations, is the detection of
far-infrared thermal emission in several QSOs at z$>$5,
tracing huge masses of dust
\citep[][]{bertoldi03b,robson04}.

\begin{figure}[t!]
\resizebox{\hsize}{!}{\includegraphics[clip=true]{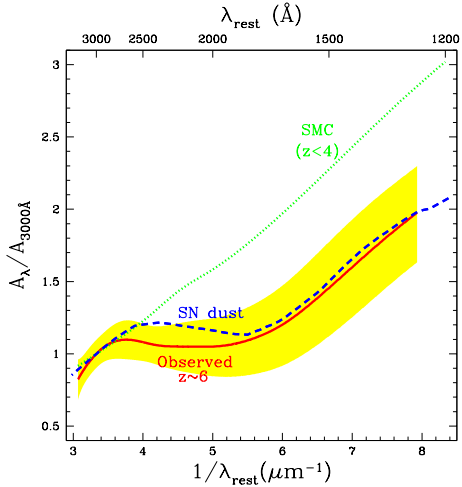}}
\caption{\footnotesize
Extinction curve observed in the QSO SDSSJ1049+46 at z=6.2 (solid curve and shaded
area) compared with the SMC curve (dotted line),
which applies to QSOs at z$<$4, and with the
extinction curve expected for dust produced by SNe
\citep[dashed line, adapted from ][]{maiolino04}
}
\label{extc}
\end{figure}

\subsection{SN dust in the early universe}\label{sn_dust}

One possible explanation is that dust is
produced in the ejecta of core-collapse supernovae, which evolve on short time
scales and therefore provide a rapid mechanism of dust enrichment.
The observational
evidence for dust production in SNe comes from the observation of SN1987A
\citep[e.g.][]{moseley89,lucy89,roche93,spyromilio93,colgan94},
and theoretically modelled by various authors
\citep[e.g.][]{todini01, nozawa03}.
The scenario of SN dust in the early universe
has been recently tested by our group
through the observation of the reddened QSO SDSSJ1048+46 at z=6.2
\citep{maiolino04}.
In particular,
we found that the extinction curve of the dust responsible for the
reddening is different with respect to that observed at z$<$4
\citep[which is SMC-like, ][]{hopkins04},
but it nicely
matches the extinction curve expected for dust produced by SNe
(Fig.~\ref{extc}). This result strongly suggests that most of the dust
in the early universe, at z$\sim$6, is produced by SNIIe. Similar results have been
obtained by \cite{hirashita05}. This investigation has
been extended to a larger number of QSOs at 4$<$z$<$6.2. Preliminary results
indicate that the extinction curve evolve gradually from the SMC-like curve
characterizing most QSOs at z$<$4, to the SN dust-dominated extinction curve at
z$=$6.2 (Maiolino et al. in prep).

\begin{figure}[t!]
\resizebox{\hsize}{!}{\includegraphics[clip=true]{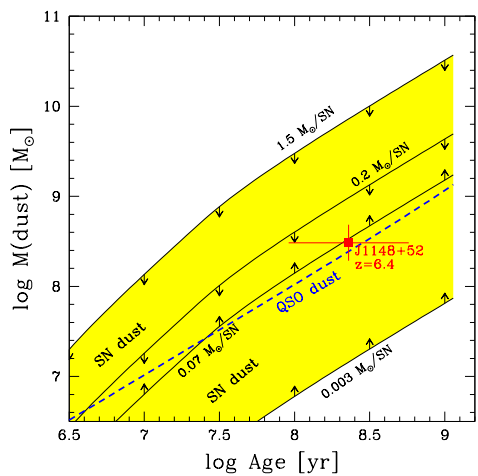}}
\caption{\footnotesize
The solid curved lines show the constraints on the
dust mass produced by SNe as a
function of age for the most distant QSO SDSSJ1148+52 at z=6.4, based on the
current observational constraints on the dust mass produced per SN (see text for
details). The dashed line is the expected
dust produced in QSO winds, according to the model of \cite{elvis02}, and under the
assumptions discussed in the text. The solid dot with errorbars indicates the
dust mass observed in
SDSSJ1148+52, assuming a formation redshift between 7.5 and 15, which appear
consistent with both the scenarios of SN-dust and QSO-dust production.
}
\label{Mdust}
\end{figure}

Recent Spitzer observations have detected only small amounts of dust in Galactic
supernova remnants (SNRs), much smaller than expected by models
\citep[][]{hines04,krause04},
thus questioning theories ascribing to SNe most of the dust
production at high redshift.
However, Spitzer observations are mostly sensitive to the warm
component of dust, which probably accounts for a minor fraction of the dust
present in SNRs. Millimetric and submillimetric data are, in principle, suited to
detect the cold component of the dust, which may account for most of the dust mass
in SNe. So far only upper limits have been reported for mm/submm cold dust
emission in SNRs, once non-thermal and foreground components are subtracted
\citep[][]{wilson05,krause04}.
Yet, Fig.~\ref{Mdust} illustrates that the current observational
constraints on the dust production in SNe are still consistent with the scenario
where most of the dust in QSOs at z$\sim$6 is produced by SNe. The curved lines
show the total dust mass produced by SNe as a function of the age of the galaxy,
assuming a star formation rate equal to that observed in the most
distant QSO at z=6.4, SDSSJ1148+52 \citep[$\rm
3000~M_{\odot}~yr^{-1}$, ][]{bertoldi03b} and a constant dust mass produced per
each supernova ($\rm M_{dust}^{SN}$). The two bottom curves show the dust
produced assuming $\rm M_{dust}^{SN}$ equal to the dust mass observed in CasA
and in the Crab Nebula through Spitzer and ISO observations
\citep[][]{hines04,green04}. Since such
observations sample only the warm component of dust and miss the cold component,
these curves should be considered as a lower limit on the total dust mass
produced. The two upper curves show upper limits on the total dust mass
assuming a $\rm M_{dust}^{SN}$ equal to different upper limits inferred for the
cold dust in CasA, based on submm data (lower curve from \citet{krause04}, upper
curve from the later value obtained by \citet{wilson05}). The solid symbol with
errorbars indicate the dust mass in the most distant QSO SDSSJ1148+52 inferred
from various mm and submm data \citep[][]{bertoldi03b,carilli04,beelen05},
including the uncertainty on the emissivity \citep[][]{dasyra05}, and assuming
a formation epoch in the range 7.5$<$z$<$15. The important result is that the
observed dust mass is consistent with the maximum dust mass production which is
allowed for SNe, based on observational constraints. However, if
the host galaxy is as young as a few times $\rm 10^7~yr$, as tentatively
suggested by \cite{walter04}, then even SNe may not have time to
produce enough dust. The uncertainties are however still very large, and more
observational data are certainly required to further investigate this issue.

\subsection{QSO--dust in the early universe}

An alternative mechanism for dust production in the early universe was proposed
by \cite{elvis02}. They show that the physical conditions (and in particular
temperature and density) in the outflowing clouds of QSOs are similar to the
conditions of AGB stellar envelopes, and therefore may provide sites of dust
formation. The interesting feature of this possible mechanism is that it is not
subject to specific time scales, and therefore it may provide a fast enrichment
mechanism in the early universe, when AGB stars had not time to evolve.
A possible caveat of this process is that it must await for the production of
metals (in particular C and Si)
from the stellar evolutionary processes. However,
we have seen in the previous section that these elements may be produced even on
relatively short time scales, and therefore this may not be a real problem
preventing the formation of dust through QSO winds in the early universe.
Unfortunately, \cite{elvis02} do not provide predictions for the extinction
curves to be compared with our observations, as in Sect.~\ref{sn_dust}.
However, it is
at least possible to check whether QSO winds can produce enough dust to account
for the dust mass observed in high-z QSOs.
We have developed a simple model by assuming
that a QSO has always been accreting at the Eddington rate, that the outflow rate
is $\rm \dot{M}_{wind}~[M_{\odot}/yr] = 0.5\times 10^{-8}~M_{BH}/M_{\odot}$
\citep[][]{proga02}, that the metallicity of the nuclear region has always
been $\rm Z=10~Z_{\odot}$, and a depletion of metals into dust as observed in
the diffuse ISM of our Galaxy. These assumption are somewhat extreme, and
therefore the results should be regarded as upper limits. However, our assumptions
are
probably not very far from reality for such rare and extreme QSOs as those
found by the SDSS at z$\sim$6.
The dashed line
in Fig.~\ref{Mdust} indicates the total dust mass produced by QSO winds under such
assumptions, and shows that the huge dust mass observed in SDSSJ1148+52 at z=6.4
can be produced even through this mechanism. Possibly, both SNe and QSO winds
may
contribute to the dust enrichment in the early universe. A more thorough analysis
will be presented in a forthcoming paper.

\begin{acknowledgements}
Limited by the space allocated to this paper, the author list could only
include a representative subsample of the people who have
been working on these projects. A more complete list of the people who
have contributed to these works includes the following:
E. Oliva, A. Ferrara, F. Mannucci,
M. Pedani, M. Roca Sogorb, F. Pieralli,
A. Beelen, F. Bertoldi, C.L. Carilli, M.J. Kaufman,
K.M. Menten, A. Omont, A. Weiss, M. Walmsley and F. Walter.
\end{acknowledgements}

\bibliographystyle{aa}

\end{document}